\documentclass[a4paper,onecolumn,12pt,fleqn]{article}
\usepackage{abstract}
\usepackage{amsmath, amsfonts, amssymb}
\usepackage{authblk}
\usepackage[english]{babel}
\usepackage[labelfont=bf,labelsep=period]{caption}
\usepackage[T1]{fontenc}
\usepackage[top=2cm, bottom=2cm, left=2cm, right=2cm]{geometry}
\usepackage{graphicx}
\usepackage{indentfirst}
\usepackage[utf8]{inputenc}
\usepackage{mathptmx}
\usepackage{microtype}
\usepackage[numbers,super,square]{natbib}
\usepackage{upgreek}
\usepackage{ragged2e}
\usepackage{setspace}
\usepackage{titlesec}
\usepackage{letltxmacro}
\usepackage{hyperref}

\providecommand{\keywords}{\raggedright Keywords:~}

\titleformat*{\section}{\normalsize\bfseries}
\titleformat*{\subsection}{\normalsize\bfseries}
\titleformat*{\subsubsection}{\normalsize\itshape}
\titleformat*{\paragraph}{\normalsize\itshape}
\titleformat*{\subparagraph}{\normalsize\itshape}
\titlelabel{\thetitle.\quad\hspace{-0.3cm}}
\titlespacing*{\section}{0pt}{*3}{*1}
\titlespacing*{\subsection}{0pt}{*2}{*1}
\titlespacing*{\subsubsection}{0pt}{*2}{*1}
\titlespacing*{\paragraph}{0pt}{*2}{*0}
\titlespacing*{\subparagraph}{0pt}{*0}{*0}

\AtBeginDocument{%
  \LetLtxMacro\hyperrefautoref\autoref
  \LetLtxMacro\autoref\firstboldautoref
}
\makeatletter
\DeclareRobustCommand\firstboldautoref{\@firstboldautoref}
\def\@firstboldautoref#1#{%
  \def\fb@autoref@star{#1}%
  \fb@autoref
}
\def\fb@autoref#1{%
  \ifcsname boldautoref@#1\endcsname
    \expandafter\hyperrefautoref\fb@autoref@star{#1}%
  \else
    \global\expandafter\let\csname boldautoref@#1\endcsname\@empty
    \textbf{\expandafter\hyperrefautoref\fb@autoref@star{#1}}%
  \fi
}
\makeatother

\title{\bfseries\raggedright\normalsize
       Record-Breaking Magnetoresistance at the Edge of a Microflake of Natural Graphite
       }

\author{Christian E. Precker,
         José Barzola-Quiquia,
        Pablo D. Esquinazi*,
               Markus Stiller,
        Mun K. Chan,
        Marcelo Jaime,
        Zhipeng Zhang,
        and Marius Grundmann
        }

\affil{C. E. Precker,  Dr. J. Barzola-Quiquia, Prof. Dr. P. D. Esquinazi, M. Stiller\\
       Division of Superconductivity and Magnetism, Felix Bloch Institute\\
       University of Leipzig, Linnéstr. 5, 04103 Leipzig, Germany.\\
       E-mail: esquin@physik.uni-leipzig.de
       }

\affil{Dr. M. K. Chan, Dr. M. Jaime\\
       National High Magnetic Field Laboratory, Los Alamos National Laboratory \\
       Pulsed Field Facility, Los Alamos, New Mexico 87545, USA.
       }

\affil{Dr. Z. Zhang, Prof. Dr. M. Grundmann \\
       Semiconductor Physics Group, Felix Bloch Institute \\
       University of Leipzig, Linnéstr. 5, 04103 Leipzig, Germany.
       }

\date{\vspace{-7ex}}

\begin{document}

\raggedright{

DOI: 10.1002/adem.201900991 \\

\textbf{Article type: Communication}

             }

\vspace{-0.5cm}

{\let\newpage\relax\maketitle}


\doublespacing
\justify

{\bf Placing several electrodes at the edge of a micrometer-size Sri Lankan natural graphite sample at distances
comparable to  the size of the internal crystalline regions, we found record values for the
change of the resistance with magnetic field. At low temperatures and at $B \sim 21$~T the magnetoresistance (MR) reaches $\sim10^7$\%.
 The MR values exceed by far all earlier reported ones for graphite and they are comparable or even  larger (at $T > 50~$K)  than the largest reported in solids including the  Weyl semimetals. The origin of this large MR lies in the existence of highly conducting 2D interfaces aligned parallel to the graphene planes.}



The electrical transport properties of bulk graphite, multigraphene and single graphene layers  show a variety of interesting phenomena. These phenomena are expected to be of advantage for  applications  such as  solar cells, supercapacitors, flexible transistors and sensors.\citep{Randviir2014,yue13} These perspectives  in addition to the high carbon abundance in nature still  attract the interest of the scientific community. In particular a detailed understanding of the electronic properties of  multigraphene samples is currently of high relevance because of the expected unique properties of the electronic band structure. We refer to   stacked graphene layers that can lead to  the formation of flat bands, i.e., a region in reciprocal space with a dispersionless  relationship, opening the possibility of triggering  high-temperature superconductivity or magnetic order.\citep{Volovik2018} This can happen  at certain localized regions of twisted graphene layers like in bilayers graphene  or at embedded interfaces between twisted Bernal or rhombohedral stacking order regions in graphite or multigraphene samples.\citep{kopninbook2015,mun13,Esquinazi2018,Cao2018a}

In this work we studied graphite samples, which are formed by stacking graphene layers held together by weak Van der Waals forces. The stacking order of the layers occurs naturally in two different ways: the hexagonal one, named Bernal with the graphene layer order ABABA… (2H), and the rhombohedral ABCABCA… (3R). Several scanning transmission electron microscope (STEM) images of the internal structure of usual graphite samples were published in the last 10 years, showing their inhomogeneity due to existence of crystalline regions of different thicknesses with different stacking orders or  twisted regions at different angles around the common $c$ axis.\citep{Esquinazi2018}

Since the 80's  and partially due to the increasing structural order and quality of the measured graphite samples, the maximum magnetoresistance (MR) found for   graphite samples steadily increased.\citep{Fauque2013} Recently published systematic studies of the MR of graphite samples of different thickness revealed that this property is directly related to the existence of two-dimensional (2D) interfaces  between crystalline regions with Bernal or rhombohedral stacking order. These 2D interfaces  are also responsible for the metalliclike behavior of the resistance of  graphite.\citep{gar12,Zoraghi2017a}  The MR of graphite samples
we discuss in this work is always measured  at fields normal to the interfaces and the graphene planes. The MR for parallel fields  is
negligible or related to a normal field component due to misalignment, which can come also from the angle distribution of the
 internal  crystalline regions (finite rocking curve width).\cite{kempa03}

Attempting to understand the nature behind the internal structure of graphite and to find a way to increase its  MR further, we have performed transport measurements under pulsed magnetic fields up to 65~T, placing the voltage electrodes on the sample edge at one side of the  sample. This enables the possibility to obtain signals to a greater extend related to the interfaces contribution, at least at  temperatures $T < 200~$K where the total conductance of the interfaces exceeds that of the semiconducting 2H and/or 3R matrix.\citep{Zoraghi2017a}

From electron back scattering diffraction (EBSD) measurements on graphite samples we know that the single crystalline regions, in the $a,b$ planes of  well-ordered bulk graphite samples are  $\lesssim 10~\upmu$m, whereas the coherent regions along the $c$ axis direction range from a few nm to several 100~nm.\citep{Esquinazi2018,arn09} The distance between grain boundaries within a single interface puts an upper limit to the typical length of the internal interfaces found in graphite samples. Therefore, in order to decrease the contribution from the grain boundaries, we need to place the voltage electrodes
at distances smaller than  $\sim 10~\upmu$m. We found that placing the voltage electrodes at the sample edge,  contacting as many interfaces edges  as possible and at small enough distances comparable to the extension of a 2D interface region, the obtained MR values are much larger than the ones reported for graphite, multigraphene or  other carbon-based materials before. The obtained MR turned out to be of the order or larger than the largest reported nowadays in solids. The technical simplification we present in this study provides a  convenient and relatively easy  way to study the response of 2D interfaces with their unconventional properties.


The electrical resistance at different positions along the sample edge was measured using four terminals. The electrodes  were made combining electron beam lithography and sputtering of Cr/Au. The aim  was to place the voltage electrodes at the edge of the thin sample in order to  directly contact   a large number of interface edges present in the sample (see, e.g., the STEM images of different graphite samples in Ref.~\cite{Esquinazi2018}). Micrometer-sized samples with well-defined interface edges are not easy to prepare. One way is to produce TEM lamellae as studied by Ballestar et al., but their production for transport measurements is very difficult, taking usually several months of preparation to get a single sample.\citep{Ballestar2013} In order to overcome these difficulties, we have developed a new method to produce graphite flakes with well-defined edges, avoiding problems of contamination or formation of an amorphous thin layer.

\begin{figure}[htb]
\includegraphics[width=\textwidth]{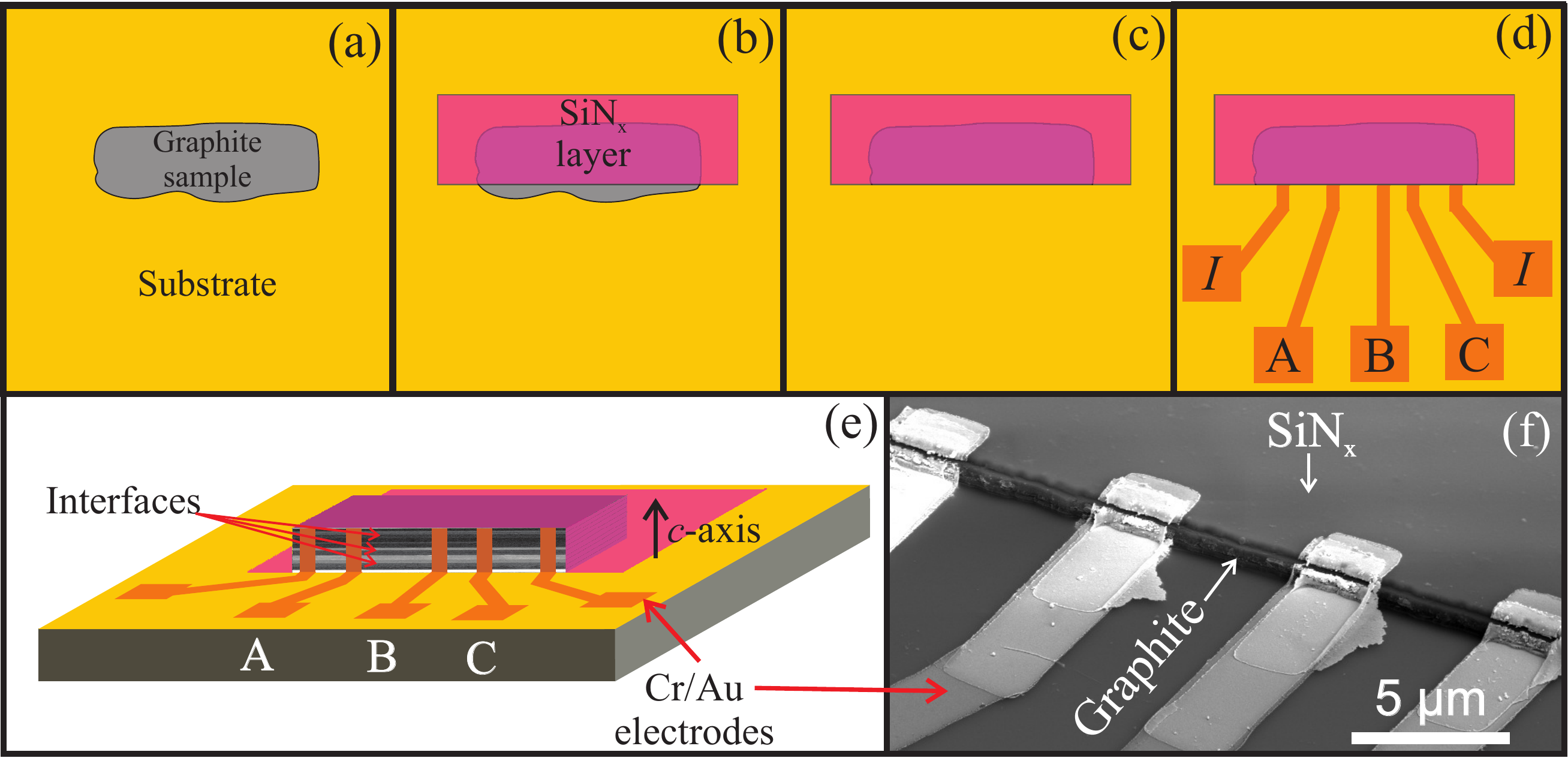}
\caption{Sample preparation sketches: (a) Mesoscopic graphite sample placed on a silicon substrate with a 150~nm $\rm Si_3N_4$ insulator layer on the top. (b) Part of the graphite sample covered with a 200 nm thick $\rm SiN_x$ layer. (c) During the RIE exposure, only the uncovered area of the sample is removed. (d) Sputtered chromium/gold electrodes were placed at different parts of the sample using electron beam lithography. The electrodes labelled  “$I$” were used to apply the electrical current through the sample and the other three electrodes were selected in pairs to measure the potential difference  at different regions of the sample, being  AB $= d_m = 5.7~\upmu$m, AC $= d_l = 13.3~\upmu$m and BC $= d_s = 4.6~\upmu$m. (e) 3D sketch of the sample with its electrodes at the edge,  parallel to the $c$ axis of the graphite structure. (f) Scanning electron microscopy image of part of the sample U11 with part of its electrodes.}
\label{Fig:sample}
\end{figure}

On the top of a $5\times5$ mm$^2$ silicon substrate with a thickness of 0.525 mm and covered with a 150 nm silicon nitride ($\rm Si_3N_4$) insulating layer, we  placed micro-flakes of Sri Lankan natural graphite (NG). The samples were from the same batch of samples analyzed with STEM, XRD and PIXE  published recently. \citep{Precker2016} After selecting flat enough samples, we covered part of the sample surfaces with a 200~nm thick $\rm SiN_x$ film using electron beam lithography (see the sketch in \autoref{Fig:sample}(a) and (b)).

An Oxford Instruments Plasma Pro NGP80 ICP device was used to etch the sample with
inductively coupled plasma (ICP) reactive ion etching (RIE). This process is very effective to remove graphene layers in graphite samples in a controlled way.\citep{Cao2018a,Yin2019,Choi2006,Prado2013} In this way the area of the sample not covered by the $\rm SiN_x$ film was completely removed, creating a sharp and well-defined edge. The area protected by $\rm SiN_x$ remained after RIE (\autoref{Fig:sample}(c)). The parameters used for RIE were 282 V for the applied DC Bias, 50~W HF power, 50~W ICP power,
$\sim 25 \times 10^{-3}~$mbar as chamber pressure, 9~sccm for Ar and 1 sccm $\rm O_2$ gas flow rate. Under these parameters, we could completely etch through a 665~nm thick graphite sample in $\sim$ 40~min.

After etching the sample, electrodes were deposited on the lateral part of the sample (see \autoref{Fig:sample}(d-f)), parallel to the $c$ axis and contacting the  interfaces edges. Electron beam lithography was used to prepare the electrodes, where a chromium thin film with thickness of 5~nm was sputtered first and then a 50~nm gold film on top. The main sample shown here, labeled U11, had in total 5 electrodes (each $\sim$ 2.5 $\upmu$m width), allowing electrical transport measurement at different regions of the sample (see \autoref{Fig:sample} (d-f)). The temperature and magnetic field dependence of the resistance to 7 T were measured in a Quantum Design $^4$He flow cryostat with a superconducting solenoid, with a high-resolution AC resistance bridge LR-700 at a frequency of 19 Hz and input current of 12 $\upmu$A.

The measurements at high magnetic fields were performed at the pulsed field facility of the NHMFL at Los Alamos National Laboratory in a $^3$He + $^4$He cryostat with maximum magnetic field of 65~T. Most of the experiments were performed with pulses of 60~T. A down-sweep pulse lasts approximately 60~ms, wherever  the up-sweep peak field is reached at $\sim$ 10 ms. An AC current of 12~$\upmu$A was applied to the sample at a frequency of 50.5~kHz ( for further details see the supporting information (SI)).
The voltages were measured with a 20 MHz sampling rate. The field was always applied normal to the graphene planes and 2D interfaces.

In order to minimize the noise on the measurement, we  used copper wires with diameter of 60 $\upmu$m, tightly twisted in pairs, with 3-4 windings per mm. We  used one pair to apply the current and two other pairs to read the voltages. The wire-pairs were glued with GE varnish on the walls of the rod used to insert the sample inside the cryostat, reducing the noise introduced by vibrations due to the pulse. Other source of noise in this kind of measurements are the open loops (untwisted parts of the wires) due to the high $dB/dt$. To minimize this effect we  fixed the twisted wires as close as possible to the sample.  Low field measurements were performed on three samples. In the main article we will show and discuss the one with the highest MR. The results of the other samples can be seen in the SI where we include further details of the samples purity.


\begin{figure}[htb]
\includegraphics[width=\textwidth]{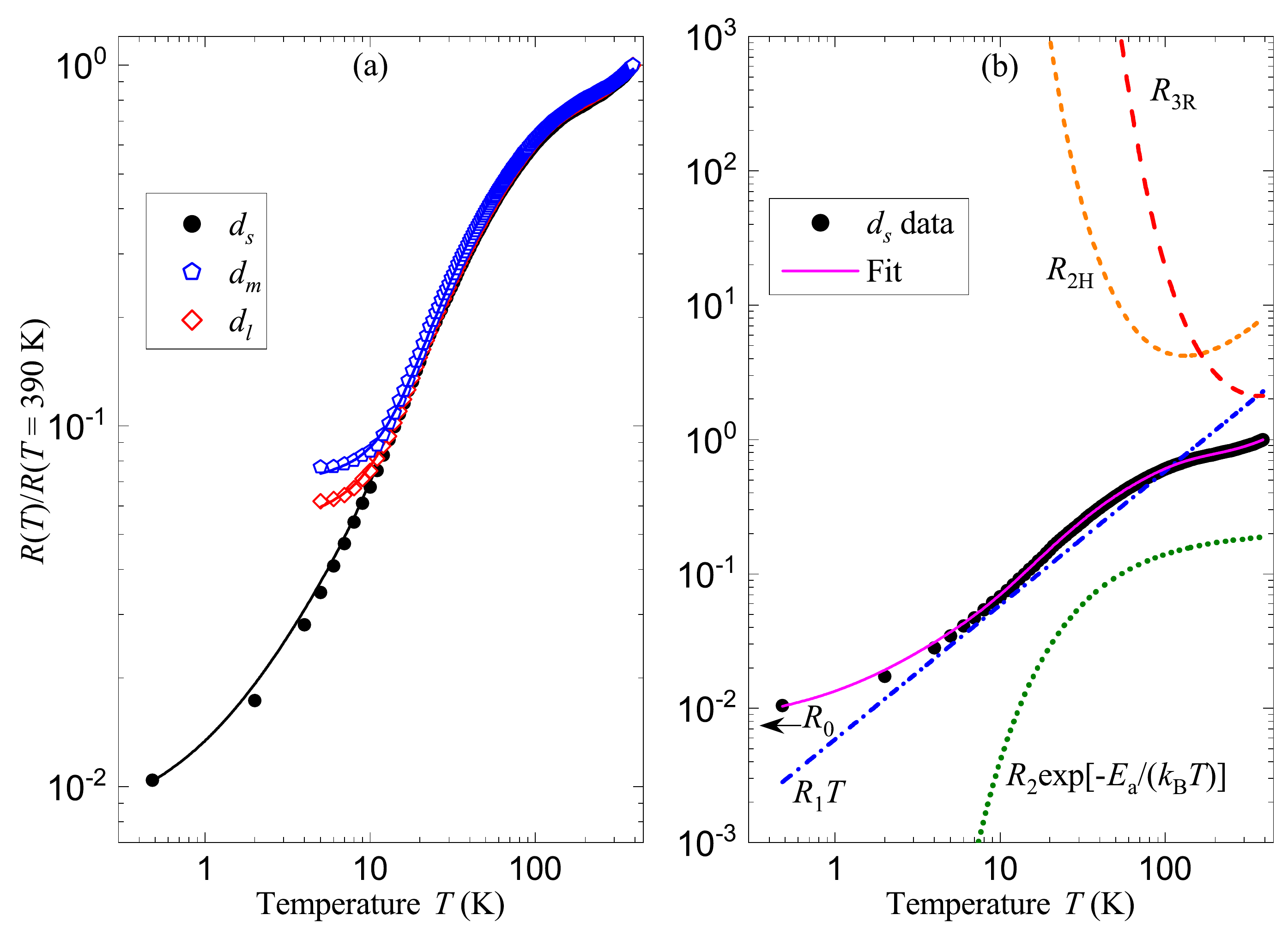}
\caption{(a) Temperature dependence of the normalized resistance
for a natural graphite sample from Sri Lanka in logarithmic scale
at different voltage-electrodes distance localized at the same
sample edge, see \autoref{Fig:sample}.  The solid lines are fits
to the phenomenological parallel resistor model, which includes
the contribution of the metalliclike 2D interfaces and the
semiconducting crystalline regions with the two stacking orders
(see Eq.~(S1) in the SI). (b) Same $d_s$ data as in (a) with the
different contributions within the parallel model.  The linear
contribution  $R_1T$ and the exponential one $R_2
\exp[-E_\text{a}/(k_\text{B}T)]$ are related to the 2D interfaces,
and $R_{2\text{H}}$ and $R_{3\text{H}}$ are related to the two
semiconducting stacking orders. The contributions were normalized
to   $R(390$~K$)$. The normalized value of the residual resistance
$R_0$ necessary to fit the low temperature data is shown by the
horizontal arrow.  $R(390$~K$) = 0.139~\Omega $. } \label{Fig:RxT}
\end{figure}

The temperature dependence of the normalized resistance at zero field for sample U11 is shown in \autoref{Fig:RxT}. The data are labeled according to the distance between the voltage reading electrodes, e.g.,  the data  labeled  $d_s$ were obtained between the electrodes B and C  with the shortest distance (see \autoref{Fig:sample}(e)). The temperature dependence shown in \autoref{Fig:RxT} follows the usual metalliclike behavior of well-ordered bulk graphite samples in all temperature range, indicating that the electrodes are sensing regions containing 2D interfaces.\citep{gar12,Zoraghi2017a,Zoraghi2018}

The results plotted in \autoref{Fig:RxT} (a) show that all three curves are similar between $T=390$~K to $\sim$ 15~K. At lower temperatures, the results labeled $d_m$ and $d_l$  tend clearly to saturate due to the contribution of a residual resistance attributed to the scattering of conduction electrons at the grain boundaries, in agreement with a large number of published data. In this temperature region and in contrast to the other two configurations, the curve $d_s$ is remarkably different exhibiting a much lower residual resistance. The resistance ratio $R$(390~K)$/R$(5~K) for $d_s$, $d_m$ and $d_l$ are 29, 13, and 16, respectively. The resistance ratio increases further for the $d_s$ configuration only, reaching a remarkable high value of $R$(390 K)$/R$(0.48 K) $\sim$ 100. The observed behavior implies that the sample is not homogeneous, in agreement with studies realized in the last years on different graphite samples.\citep{Ballestar2013,Zoraghi2018}

The temperature dependence of the resistance is fitted using a phenomenological parallel resistor model as proposed first in Ref.~\cite{gar12} and extended in following years.\citep{Zoraghi2017a,Zoraghi2018} The model takes explicitly into account  the internal structure of real graphite samples,  assuming  three contributions in parallel. The first one due to the embedded 2D interfaces provides the metalliclike behavior of graphite; the second and third ones are the semiconducting contributions of the hexagonal 2H and rhombohedral 3R stacking orders, see the SI for details of the model. The metalliclike contribution is composed by a temperature independent residual resistance, a linear and a thermally activated temperature dependent term. Such phenomenological model describes with good accuracy the temperature dependence of the resistance in the entire investigated  temperature range. Note that below $\sim$ 200~K, the interface contribution is the most important one and the fit is not very sensitive to the parameters of the other two contributions, see \autoref{Fig:RxT}(b). A detailed discussion on this issue and on the weight of the parameters in a given temperature range was published recently.\citep{Zoraghi2017a} The values of the fit parameters can be seen in the SI. As expected, the residual resistance from the fit at the low temperature of the $d_s$ data shown in \autoref{Fig:RxT}(a) is one order of magnitude smaller than for $d_m$ and $d_l$. From the fits of the data to the parallel resistor model we find that the linear-in-temperature term of the interface contribution\citep{Zoraghi2017a} is important at $T < 10$~K, whereas the thermally activated exponential term (with an excitation energy of the order of $\sim 5$~meV) clearly contributes between 15~K and $\sim 200$~K. It is interesting to note that the low residual resistance of the $d_s$ data clearly reveals  the  linear-in-temperature contribution that holds to the lowest measured temperature.
The temperature dependence of the resistance at different constant applied magnetic fields for the three configurations of sample U11, the  field-driven metal-insulator transition and the Shubnikov-de Haas (SdH) oscillations are discussed in the SI.

\begin{figure}[htb]
\includegraphics[width=\textwidth]{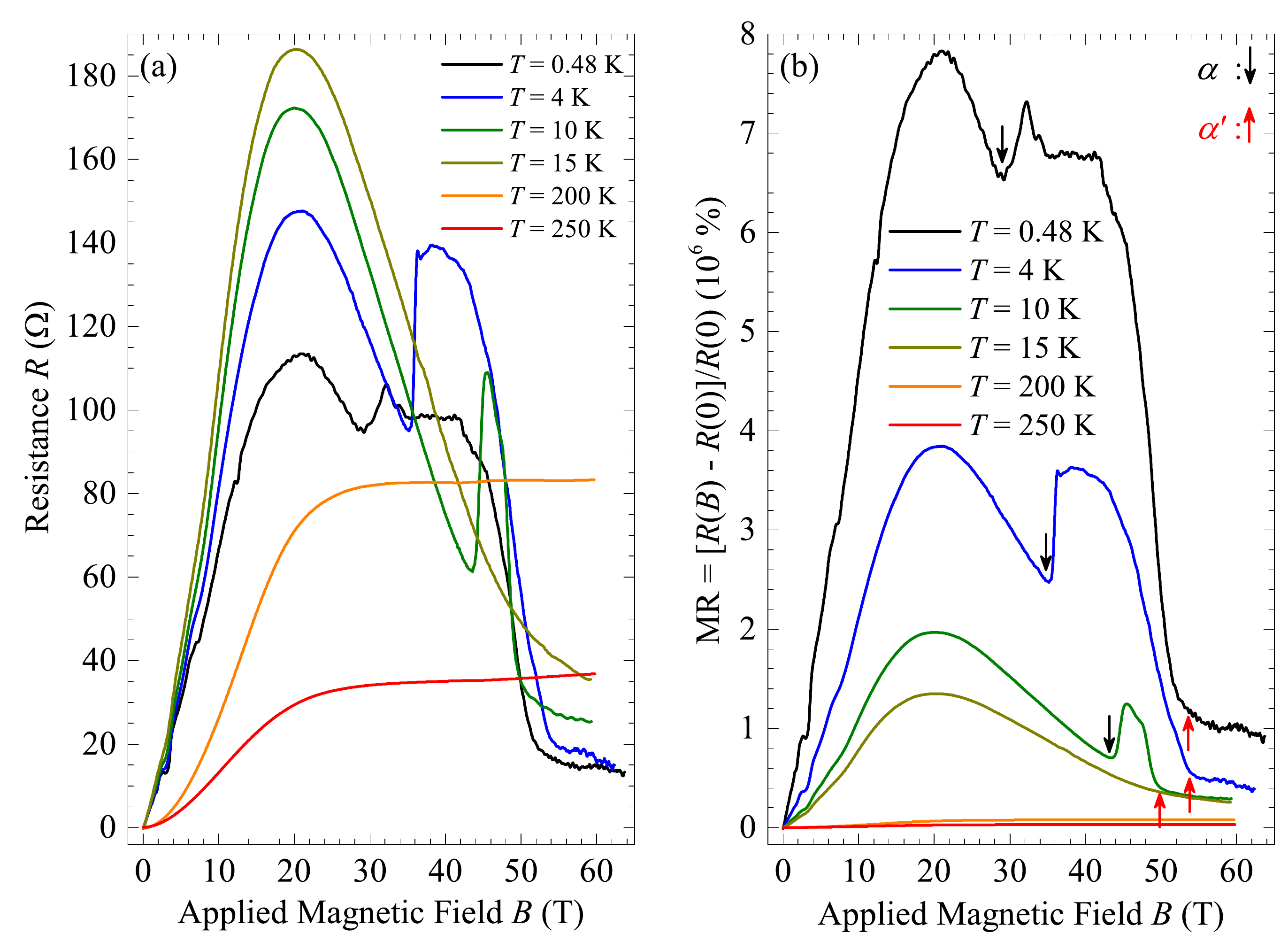}
\caption{Magnetoresistance measured at $d_s$ as a function of
magnetic field  at different constant temperatures. (a) Absolute
value of the resistance and (b) the normalized magnetoresistance
(in units of $10^6\%$) vs. applied field. The black-down arrows
indicate the onset transition $\alpha$ and the red-up arrows the
so-called re-entrant transition $\alpha'$.} \label{Fig:MRsample}
\end{figure}

\begin{figure}[htb]
\includegraphics[width=\textwidth]{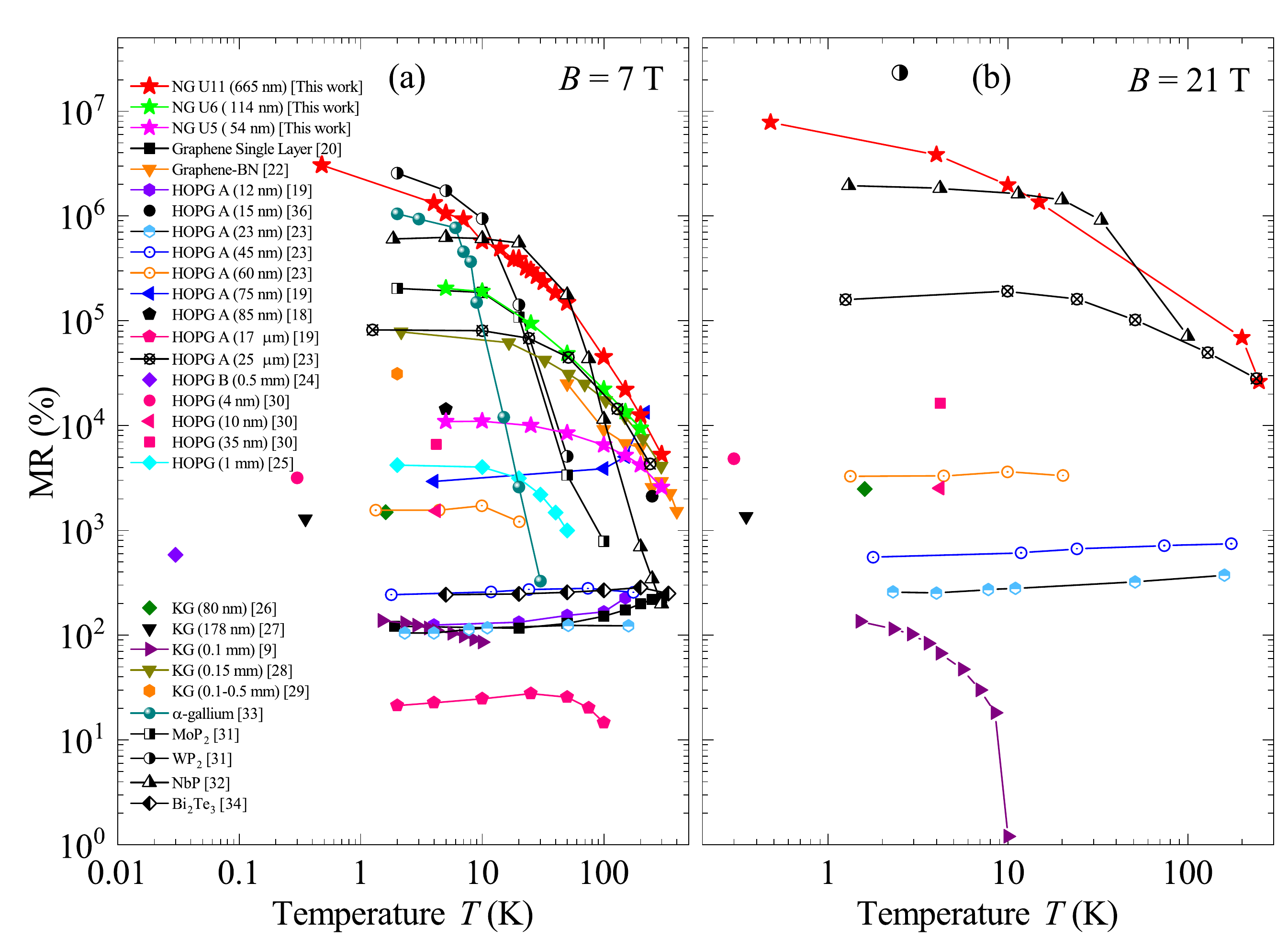}
\caption{Magnetoresistance of different graphite samples from
literature,  three Weyl semimetals,  $\alpha$-gallium and
topological insulator Bi$_2$Te$_3$. The legend shows  the kind of
sample, where HOPG stands for "highly oriented pyrolytic
graphite", and its grade (A or B), KG means "Kish graphite" and NG
"natural graphite". The number in brackets represents the sample
thickness given in the corresponding publication.  (a) Shows the
temperature dependence of the MR at an applied field of $B = 7$ T
and (b) at $B = 21$ T.} \label{Fig:MRliterature}
\end{figure}

We discuss now the field dependence of the resistance and the  MR defined as
MR $=[(R(B) - R(0)] / R(0)$, where $R(0)$ is the measured resistance at zero applied magnetic
field (MR $(\%) = 100\%  \times =[(R(B) - R(0)] / R(0)$). As shown in the SI, and as example, at $T=5$ K and $B=7$~T  the MR reaches $5 \times 10^5$ \%  at the contacts $d_m$ and $d_l$.  For the contact configuration $d_s$ the MR is nearly twice larger reaching $\sim 10^6$ \%, see  \autoref{Fig:MRsample} for the  results obtained with pulsed fields. Earlier studies in high quality graphite samples show a MR of $\sim$ 15,500 \%, 14,000 \%, 400 \%, and 75 \%.\citep{Zoraghi2018,Barzola-Quiquia2008,Gopinadhan2013,Zhang2004} In graphene/boron-nitride heterostructures a MR of $\sim$ 90,000 \% was measured at similar field and temperature\citep{Gopinadhan2015}, see  \autoref{Fig:MRliterature}(a).

\autoref{Fig:MRsample} shows the high field results obtained in the $d_s$ configuration; panel (a)  shows the absolute resistance  and  panel (b) the MR at different constant temperatures. The up and down arrows in (b) indicate the fields at which the commonly reported electronic high field transitions $\alpha$ and $\alpha'$ of graphite occur at $T < 15~$K. These transitions as well as the maximum and negative MR above 20~T are related to the electronic 2D systems of some interfaces in the sample and were discussed in detail in a recent publication. \citep{bar19} The MR at $T = 0.48$ K and $B \sim 21$~T reaches $\sim 8 \times 10^6$ \%, exceeding by far all values reported for graphite in literature.

A comparison with the temperature dependence of the MR data  reported for different graphite samples in the literature and at fields of 7~T and 21~T, is given in \autoref{Fig:MRliterature}(a) and (b).\citep{Hubbard2011,Kopelevich2010,Taen2018,Taen2018a,Kopelevich2003a,kempa02,bru18}
 Large MR values were observed for Type-II Weyl semimetal like $\rm WP_2$\citep{Kumar2017} reaching MR $\sim 2 \times 10^6$ \% at low temperatures and at 7~T, similar to the MR we obtained at  $d_s$,  see \autoref{Fig:MRliterature}(a). Further data of the semimetals MoP$_2$ \citep{Kumar2017} and NbP \citep{she15},  the metallic  sample $\alpha$-gallium\citep{Chen2018} and of the topological insulator Bi$_2$Te$_3$ \cite{wan12bite} are shown in \autoref{Fig:MRliterature}(a,b).
We note that the MR of graphite at both fields and at the configuration $d_s$ reaches values comparable or even larger (at $T > 50~$K ) than the largest so far reported.

\paragraph{Possible origin of the huge MR measured in graphite:}~ First, we note that the increase of the MR  decreasing the distance between the voltage electrodes (the voltage electrodes distance in the configuration $d_s \simeq d_l/3$) is not related to an increase in the contribution of a ballistic transport. Measurements of the MR in thin graphite samples with no or a low number of interfaces showed that the MR is not only much smaller but {\em decreases with the sample size} due to the large mean free path and huge mobility of the carriers within the semiconducting graphene layers in the graphite matrix.\citep{gon07,dus11} Experimental studies  clearly showed that the large MR as well as the SdH oscillations of graphite are directly related to the response of the 2D interfaces, i.e., they are not intrinsic of the ideal graphite structure and not related to the intrinsic carriers within the graphene layers.\citep{Zoraghi2018,bar19}  The value of the MR of graphite samples depend  on the thickness of the sample, as one recognizes in the results of the measured samples in this work, see \autoref{Fig:MRliterature}(a), in agreement with Refs.\cite{bar19,bru18}.  For small enough sample thickness the number of interfaces decreases and several features of the MR vanish. For example, the maximum at $\sim 20$~T, the negative MR and the electronic phase transitions observed at  low enough temperatures and above $\sim 20$~T, completely vanish. \citep{bru18,bar19} The semiconductinglike behavior observed in thick samples at high temperatures, see the curves at 200~K and 250~K in \autoref{Fig:MRsample}, or even at lower temperatures in much thinner samples,\citep{bru18} can semiquantitatively be explained with a semiconducting two-band model.\citep{bar19}

These facts plus the giant magnetic anisotropy clearly indicate that the origin of the huge MR has to be found within the 2D electronic system at the interfaces embedded in the graphite samples. Taking into account that the 3R stacking order remains a minority phase in our samples (less than 15~\%), the interfaces between twisted 2H stacking regions (type I) and between 2H and 3R regions (type III) are the most probable ones. The possible occurrence of superconductivity at these 2D interfaces has been shown experimentally\citep{Cao2018a,Esquinazi2018,Ballestar2013,Precker2016,bal15,sti18} and theoretically predicted.\citep{Volovik2018} The origin of the thermally activated exponential increase with temperature, one part of the interface contribution to the total resistance (see  \autoref{Fig:RxT}(b)),   remains still controversial.\citep{Zoraghi2017a} We note, however, that it has been already observed in superconducting thin films, granular superconductors and in artificially grown Josephson-junction arrays.\citep{mas99,sha83,zan96} For a discussion of the effects of granular superconductivity on the MR  of the 2D interfaces of graphite we refer to a recently published study.\citep{bar19}
We suggest, therefore, that at least part of the observed large MR
can be related to the existence of granular superconductivity at certain 2D interfaces embedded in the graphite matrix.\citep{Esquinazi2018}
Further increase in the MR of graphite samples can be achieved by reducing the residual resistance measured in series with the interface resistance. This should be possible through the reduction of the electrode distance or trying to contact an interface where superconductivity is less granular. Obviously, in this case the MR would diverge.

Finally, we would like to compare  the MR we obtain in graphite with that of the Weyl semimetals.  In  \autoref{Fig:MRliterature} one recognizes that  the three reported MR of Weyl semimetals and
that of $\alpha$-gallium shows a similar behavior with temperature:
below a sample-dependent temperature it tends
to saturate, whereas above this temperature it  decreases with temperature much more steeply  compared to  the MR of the graphite samples.
This result plus the fact that the MR of graphite thick samples is mainly related to the electronic
systems at the 2D interfaces, already suggest that the origin as well as the mechanisms involved in the MR are not the same,
in spite of the expected similarities in the band structure.
Moreover, due to the parallel contributions of different electronic systems in the graphite samples, the observed decrease with temperature of the MR  is partially related to the relative increase of the conductance of the semiconducting regions, see \autoref{Fig:RxT}(b), which show a much smaller MR than that of the 2D interfaces. Therefore, we  expect that a weaker decrease of the MR
with temperature can be achieved reducing the parallel contribution of the semiconducting regions around the interfaces.

In conclusion, with voltage electrodes separated by few micrometers along the edge of graphite
samples we   measured the longitudinal resistance  at different  regions of  the sample edge.
This experimental method  enables the study of the transport properties of interfaces embedded in the graphite matrix.
The obtained results indicate that graphite is an inhomogeneous material at a scale of a few micrometers within the $a,b$ planes.
This inhomogeneity is one main factor that affects substantially the measured magnetoresistance at low  temperatures.
 The MR of the graphite interfaces is very large, exceeding in some temperature and field region  the largest MR values  reported for solids.

\section*{Acknowledgments}

C.E.P. gratefully acknowledges the support provided by the Brazilian National Council for the Improvement of Higher Education (CAPES) under 99999.013188/2013-05. The studies were supported by the DAAD Nr. 57207627 ("Untersuchungen von Grenzfl\"achen in Graphit bei sehr hohen Feldern") and partially supported by the DFG under ES 86/29-1 and the Collaborative Research Center (SFB762)
  ``Functionality of Oxide Interfaces'' (DFG). Work at the National High Magnetic Field Laboratory was supported by the US Department of Energy, the State of Florida, and the National Science Foundation through cooperative grant agreements DMR 1157490 and DMR 1644779. M.J.
  and M.K.C. acknowledge the support by the  US Department of Energy BES ``Science at 100 T" grant, Division of Materials Science and
  Engineering.

\section*{Authors contributions}
C.E.P. and J.B.-Q. were responsible for the samples preparation.
Z.Z. was responsible for SiN$_x$ deposition. M.S. was responsible
for AFM measurements. M.K.C., M.J., J.B.-Q. and C.E.P. conducted.
the high field measurements. C.E.P., P.D.E. and J.B.-Q. analyzed
the data, whereas P.D.E. and J.B.-Q. contributed equally. M.G.
gave the idea to make use of RIE in order to measure the
interfaces. P.D.E. conceived the experiment(s), and took the lead
writing of the manuscript. All authors provided critical feedback
and helped to shape the research, analysis and manuscript.

\bibliographystyle{advmatcep}

\vspace{1cm}

\raggedright{{\Large Supporting Information}}


{\let\newpage\relax\maketitle}


\section*{Magnetic impurities characterization}

The magnetic impurities concentration of the samples was analyzed
by RBS/PIXE measurements using 2~MeV proton beam with diameter of
0.8~mm and a penetration depth of $\sim 35~\mu$m, for more details
see Refs. \cite{pre16,chap3}. The measurements (magnetic analysis
as well as magnetoresistance) were always performed in an
exfoliated graphite surface, because the surface of the samples
can be contaminated, see, e.g. \cite{spe14}. The analysis
indicates a concentration of 6.4~ppm of Fe and 5.9~ppm of Ti (ppm
means $\mu$g element per gram sample). The concentration of other
elements was below the detection limit of $\sim 1$~ppm. Clearly,
the small amount of impurities is not the reason for the huge
magnetoresistance. Furthermore, we note that magnetic impurities
or magnetic order through defects originate a (small) negative,
not positive, magnetoresistance in graphite \cite{chap3}.

\section*{Similarities between D.C. and pulsed fields measurements}
As a pre-characterization we have measured the magnetoresistance
of the same samples to 7~T D.C. fields and using  an input current
frequency of 19~Hz at similar temperatures. The results indicate
that there are no differences between the absolute resistance and
magnetoresistance between the D.C. and pulse fields measurements.
A direct comparison can be obtained from the data  shown in
Ref.~\cite{bar19} and the SI included in that publication.

Concerning the time scale of the magnetic pulse: the up-sweep from
0 to 60 T took  $\sim 10$~ms, whereas the down-sweep from 60 T to
0~T  $\sim 60$~ms. Therefore, there was a significant time
difference in the characteristic experimental time scales for up
and down sweeps. The measurements at 50.5~kHz input current
frequency means that
 the average field difference between measured points were 0.12~T for up-sweeps and 0.02~T  for down-sweeps.
 Qualitatively and quantitatively  both measurements were similar.
 To a certain extent, this  corresponds also to an in-situ test of the frequency independence of the shown results.

\section*{Sample U11}
\subsection*{Temperature dependence of the resistance}

The fits in Figure 2 of the main article were obtained using a
phenomenological  model with an equivalent total resistance $R_T$
composed by 3 contributions in parallel taking into account the
internal structure of the graphite samples. The first one
corresponds to  the interfaces $R_\textit{i}$; the second one to
the crystalline region of the hexagonal stacking order
$R_\text{2H}$, and the third one,  the  crystalline region of the
rhombohedral stacking order $R_\text{3R}$, given by
Eq.~(\ref{Eq:Rtotal}).\citep{Garcia2012,Zoraghi2017}

\begin{equation}
    \frac{1}{R_T(T)} = \frac{1}{R_i(T)} + \frac{1}{R_{\rm 2H}(T)} + \frac{1}{R_{\rm 3R}(T)}\text{,}
    \label{Eq:Rtotal}
\end{equation}
where the resistance contribution of the 2D interfaces is assumed
to be of the form
\begin{equation}
    R_i(T) = R_0 + R_1 \cdot T + R_2 \cdot \exp\left(\frac{-E_\text{a}}{k_\text{B} \cdot T}\right)\text{,}
    \label{Eq:Rinterface}
\end{equation}

with $R_0$ a temperature independent factor that represents the
residual resistance. As written in the main article the origin of
the  thermally activated contribution remains still controversial;
we speculate that its origin is related to thermally activated
behavior between superconducting regions localized at the 2D
interfaces (see Refs. in the main article). $R_1$, $R_2$, and the
activation energy $E_\text{a}$ are free parameters and
$k_\text{B}$ the Boltzmann constant. The crystalline regions where
the 2D interfaces are embedded show a semiconductinglike behavior
of the form:
\begin{equation}
    R_n(T) = a_n \cdot T^{3/2} \cdot \exp\left(\frac{-E_{gn}}{2k_B \cdot T}\right)\text{,}
    \label{Eq:Rcrystals}
\end{equation}
where $n$ = 2H for the hexagonal contribution and $n$ = 3R for
rhombohedral. $a_n \cdot T^{3/2}$ is a mobility pre-factor, which
depends on parameters like the mean-free-path and the carrier band
structure, the parameter $a_n$ and the gap energy $E_{gn}$ are
free parameters. \autoref{tab:parameters} shows the parameters
used in Eq~(\ref{Eq:Rtotal}) to fit the experimental data in
Fig.~2 of the main article.

\begin{table}[!htb]
\caption{Normalized parameters obtained from the fits to
Eq.~(\ref{Eq:Rtotal}) of the temperature dependence of the
resistance of sample U11. The normalization value  of $R_i$ and
$a_i$ is $R(390$K$) = 0.139~\Omega$.} \centering
\begin{tabular}{l @{\hspace{30pt}} c @{\hspace{18pt}} c @{\hspace{18pt}} c}
\hline\hline & $d_s$ & $d_m$ & $d_l$ \\ \hline
Electrodes distance $\left(\upmu\text{m}\right)$ & 4.6 & 5.7 & 13.3 \\
$R_0~\left(10^{-2}~\right)$ & 0.7543 & 6.504 & 4.712 \\
$R_1~\left(10^{-3}~\text{K}^{-1}\right)$ & 5.848 & 1.746 & 2.484 \\
$R_2~\left(10^{-1}~\right)$ & 2.073 & 6.780 & 5.964 \\
$E_\text{a}~\left(\text{meV}\right)$ & 3.396 & 4.281 & 4.045 \\
$a_{\text{2H}}\left(10^{-3}~\text{K}^{-3/2}\right)$ & 0.6305 & 20.18 & 2.447 \\
$E_{g\text{2H}}~\left(\text{meV}\right)$ & 33.68 & 20.00 & 23.21 \\
$a_{\text{3R}}~\left(10^{-5}~\text{K}^{-3/2}\right)$ & 6.367 & 10.07 & 9.430 \\
$E_{g\text{3R}}~\left(\text{meV}\right)$ & 98.12 & 103.8 & 100.7 \\
\hline\hline
\end{tabular}
\label{tab:parameters}
\end{table}

\begin{figure}[!htb]
\includegraphics[width=0.8\textwidth]{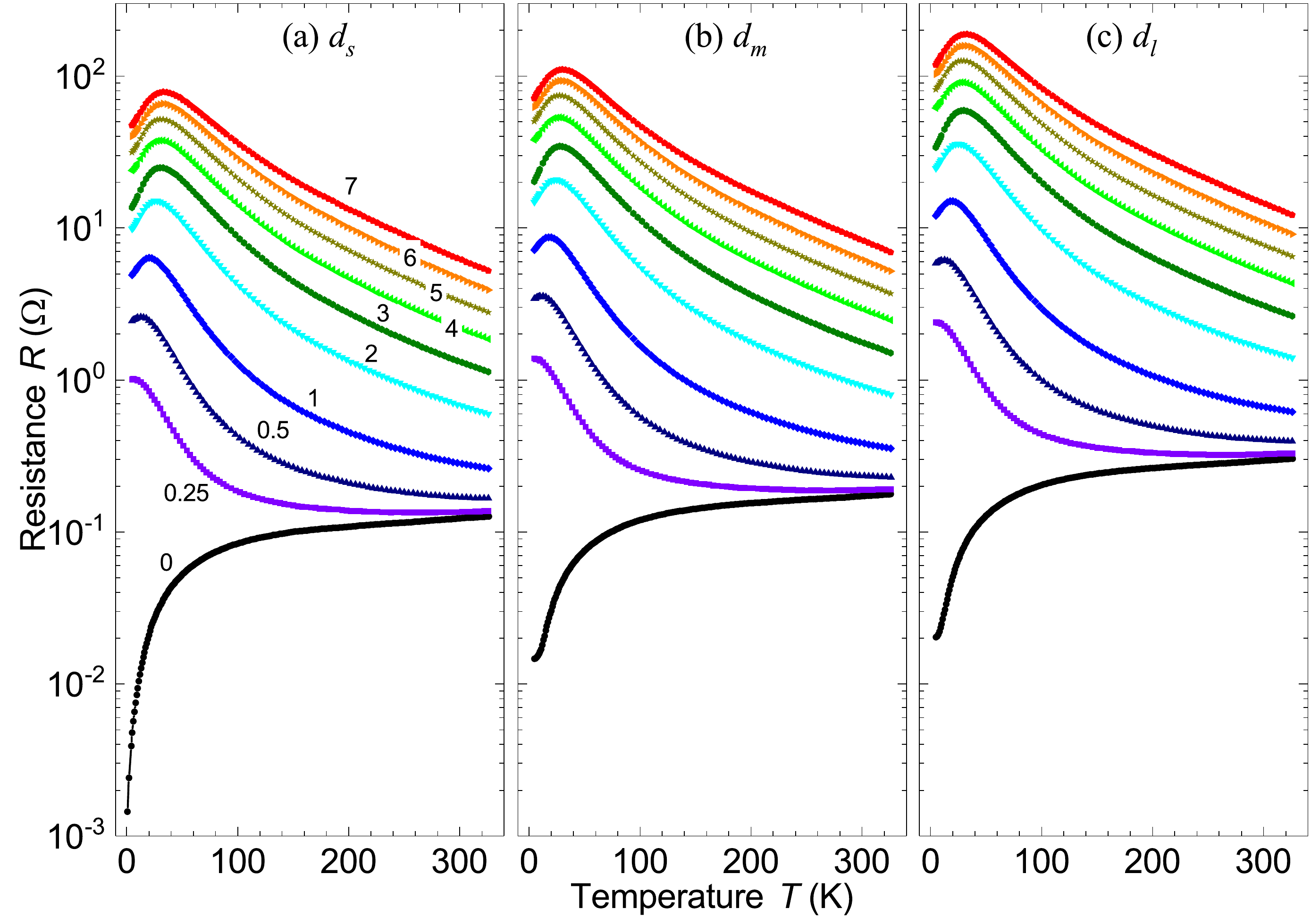}
\centering \caption{Temperature dependence of the longitudinal
resistance of natural graphite at different magnetic fields
applied normal to the internal interfaces and graphene planes. (a)
Resistance measured at the $d_s$ electrodes, (b) at the $d_m$
electrodes, and (c) at the $d_l$ electrodes. The x- and y-scale
are the same for all graphs. The numbers at the curves in (a)
represent the applied magnetic field in T. The same labels/curves
colors apply for the curves in (b) and (c).} \label{fig:Tdep}
\end{figure}

The temperature dependence of the resistance at different constant
applied magnetic fields for all three configurations of sample U11
is presented in \autoref{fig:Tdep}. The typical field-driven
metal-insulator transition (MIT) of graphite samples with
interfaces is clearly observed .

\subsection*{Magnetoresistance at DC magnetic fields to $\pm 7$T}

\begin{figure}[!htb]
\includegraphics[width=0.8\textwidth]{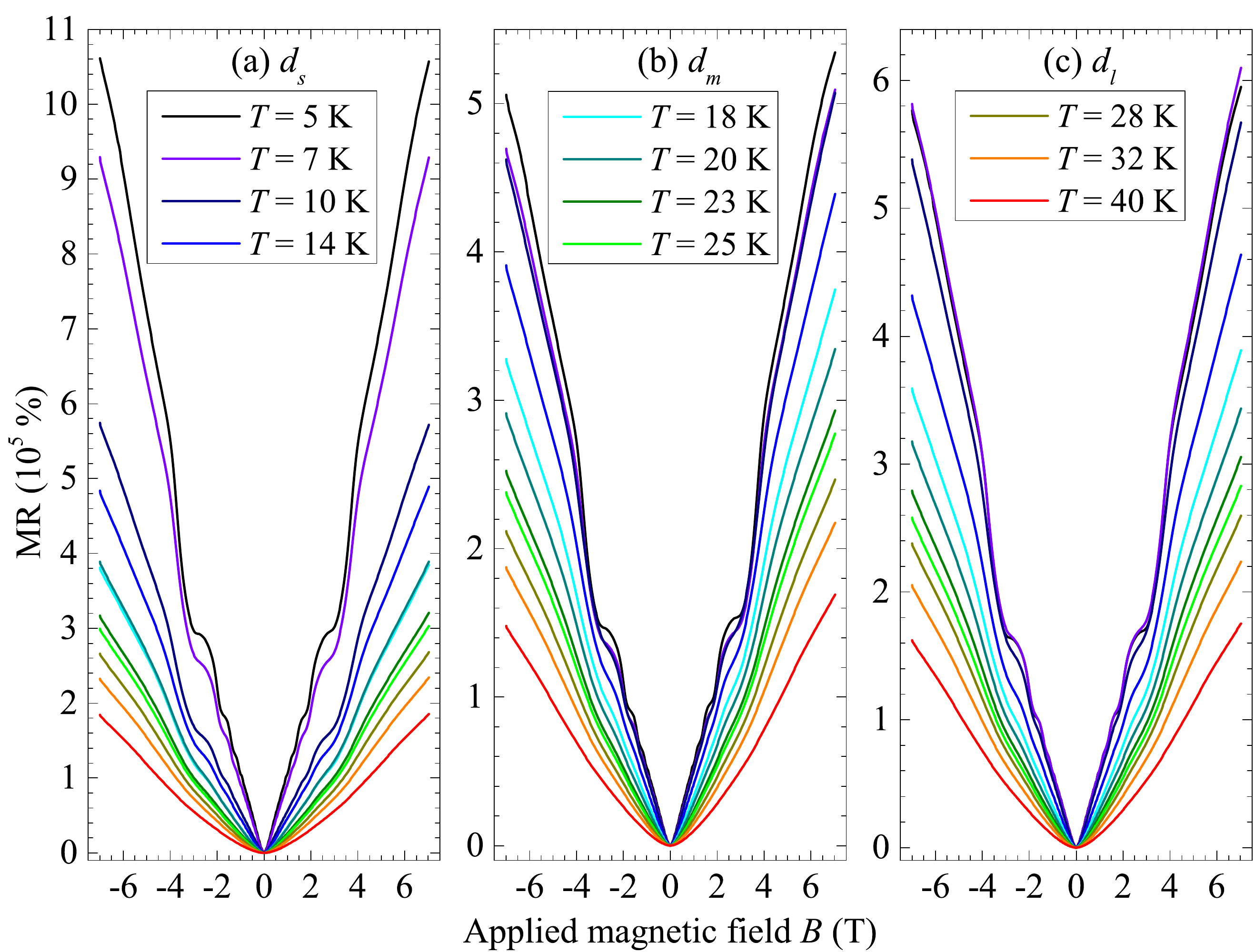}
\caption{Magnetoresistance defined as MR $= {\lbrace[R(B) -
R(0)]/R(0)\rbrace \times 100~\%}$, at different constant
temperatures (curve-colors/temperature apply to all graphs) with a
maximum  field of $B = \pm 7$~T applied perpendicular to the 2D
interfaces and graphene layers.} \label{fig:MR7T}
\end{figure}

\autoref{fig:MR7T} shows the magnetoresistance (MR) at all
contacts combinations, $d_s$, $d_m$, and $d_l$ at different
temperatures. The quantum oscillations, known as Shubnikov-de Haas
(SdH) oscillations, are visible at lower temperatures. The SdH
oscillations are a signature of the presence of 2D interfaces on a
graphite sample.\cite{Zoraghi2018} Graphite samples without the 2D
interfaces show no SdH oscillations. The reason is that the
semiconducting crystalline regions have an exponentially small
carrier concentration at low temperatures.

\begin{figure}[!htb]
\includegraphics[width=0.9\textwidth]{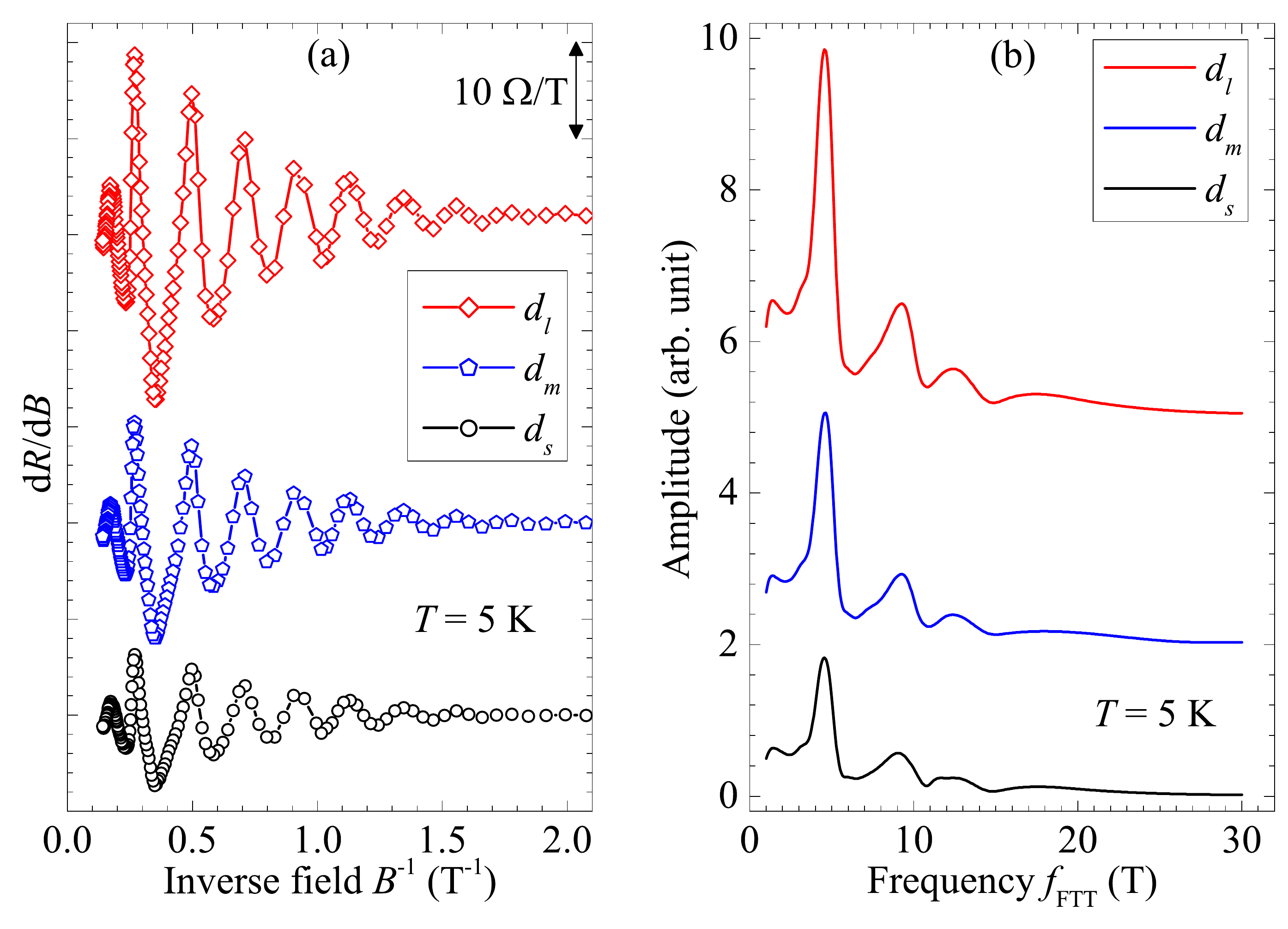}
\caption{(a) First derivative of the longitudinal resistance $R$
versus the inverse of the
         applied magnetic field $B$ at each electrode pair at $T=5$ K. (b) Fourier transform
         of the data in (a).}
\label{fig:SdH-FFT}
\end{figure}

\autoref{fig:SdH-FFT} (a) shows the first derivative of the data
presented in \autoref{fig:MR7T} at $T=5$~K.  One can see that the
period of the oscillations does not appreciably change  with the
electrodes locations, only the amplitude of the oscillations
changes. For an interpretation of the SdH oscillations in terms of
the 2D interfaces and the meaning of the 2D carrier density one
obtains from the SdH oscillation frequency we refer to  recently
published studies.\cite{Zoraghi2018}
\subsection*{Magnetoresistance at pulsed magnetic fields to 65~T}
\begin{figure}[!htb]
\includegraphics[width=0.9\textwidth]{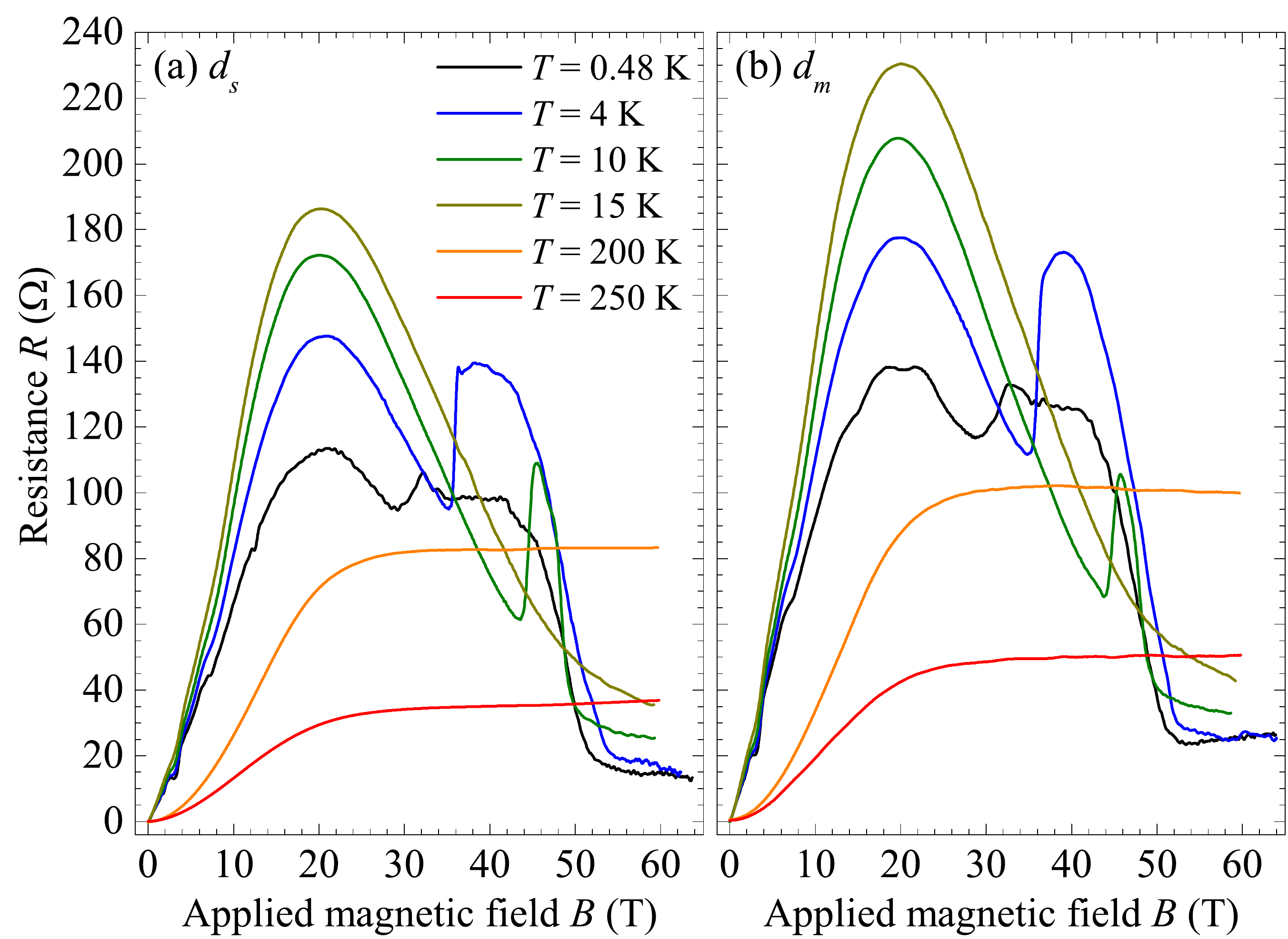}
\caption{Resistance measured simultaneously at the two
electrode-pair configurations (a) $d_s$ and (b) $d_m$, at constant
temperatures. The y- and x-axis scales are the same  in (a) and
(b).} \label{fig:HF-ds-dm}
\end{figure}

The results of the pulsed  magnetic fields  applied parallel to
the $c$ axis at different constant temperatures are shown in
\autoref{fig:HF-ds-dm}. Due to difficulties in contacting the
sample to perform the measurement simultaneously in all three
configurations, we have measured only  $d_s$ and $d_m$.
\autoref{fig:HF-ds-dm} (a) and (b) have the same x- and y-axis
scale and they share the same legend. The data showed in
\autoref{fig:HF-ds-dm} was measured simultaneously at each
temperature, e.g., one single pulse was used to store the data of
the resistance at $d_s$ and  $d_m$.

At $T = 250$~K  and 200~K the resistance shows a high increase up
to $B \sim$ 25~T and  a weaker increase at higher fields. At  $T
\leq 15$~K, the resistance reaches a maximum at $B \simeq$ 21~T
and a negative MR at higher fields. Explanations for the change of
slope of the MR from positive to negative and the electronic
transitions can be found in Ref.~\cite{bar19}.

Besides sample U11 we have produced and measured  samples U5 and
U6 from the same batch of natural graphite from Sri Lanka, using
the same technique and placing the contact electrodes also at the
edge of these samples. They were characterized at low fields up to
$B=7$ T. Measurements at higher field  could not be performed
because the electrodes burned just before the measurements.

\section*{Sample U5}

\begin{figure}[!htb]
\includegraphics[width=0.8\textwidth]{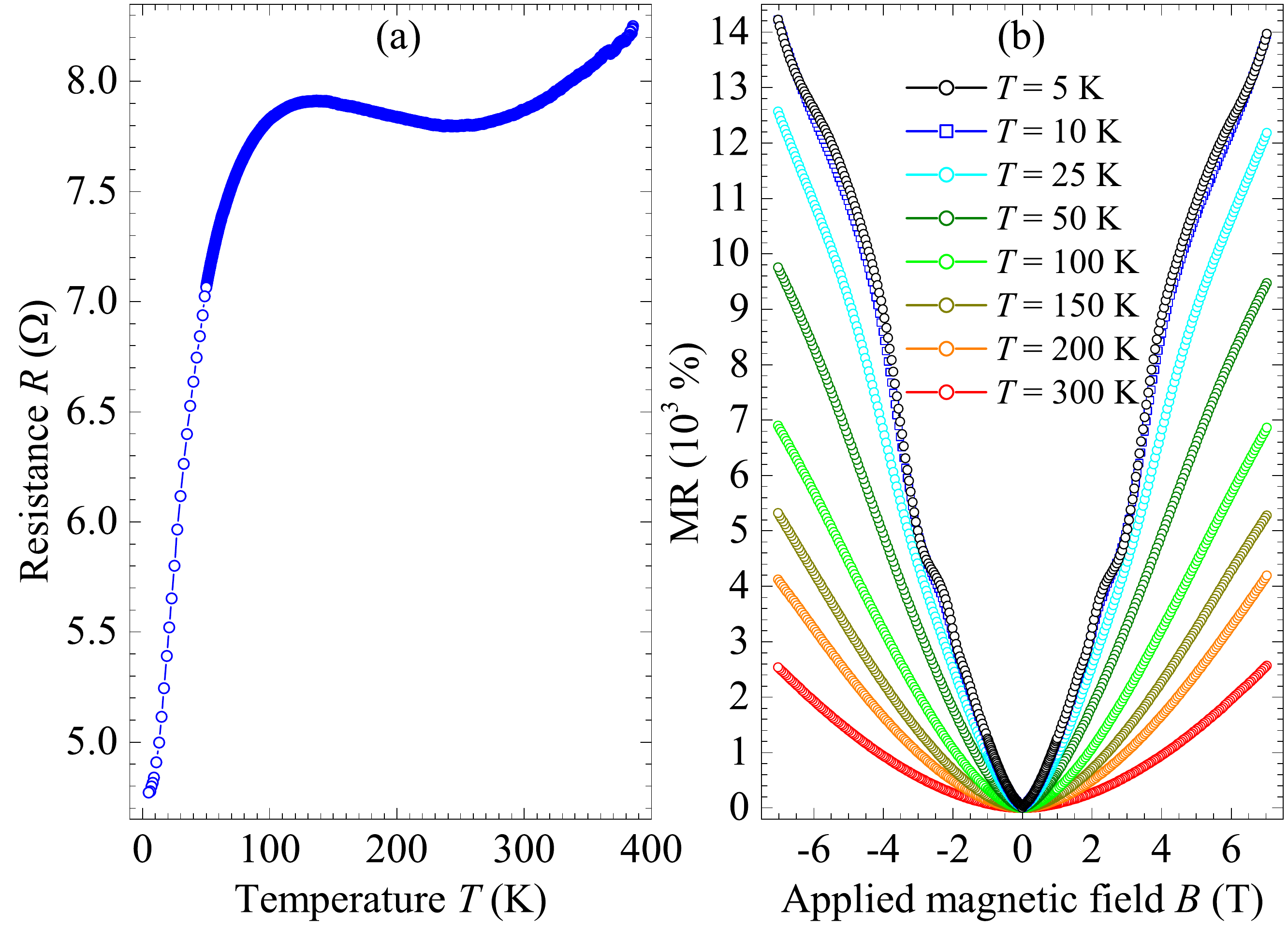}
\caption{Data of sample U5: (a) temperature dependence of the
resistance. (b) The magnetoresistance
         to $B=\pm 7$~T at different temperatures.}
\label{fig:SampleU5}
\end{figure}

\autoref{fig:SampleU5}(a) shows the temperature dependence of
sample U5 of 54~nm thickness. This sample shows a metalliclike
behavior from $T=390$~K to  $\sim$ 250 K. At lower temperatures to
$\sim$ 125 K the resistance increases below $\sim 100$~K  a
re-entrance to the metalliclike behavior is observed with a
resistance ratio of $R(T = 300 ~\text{K})/R(T = 5 ~\text{K})
\sim~1.65$. \autoref{fig:SampleU5}(b) shows the  MR at different
constant temperatures.

\section*{Sample U6}

\begin{figure}[!htb]
\includegraphics[width=0.8\textwidth]{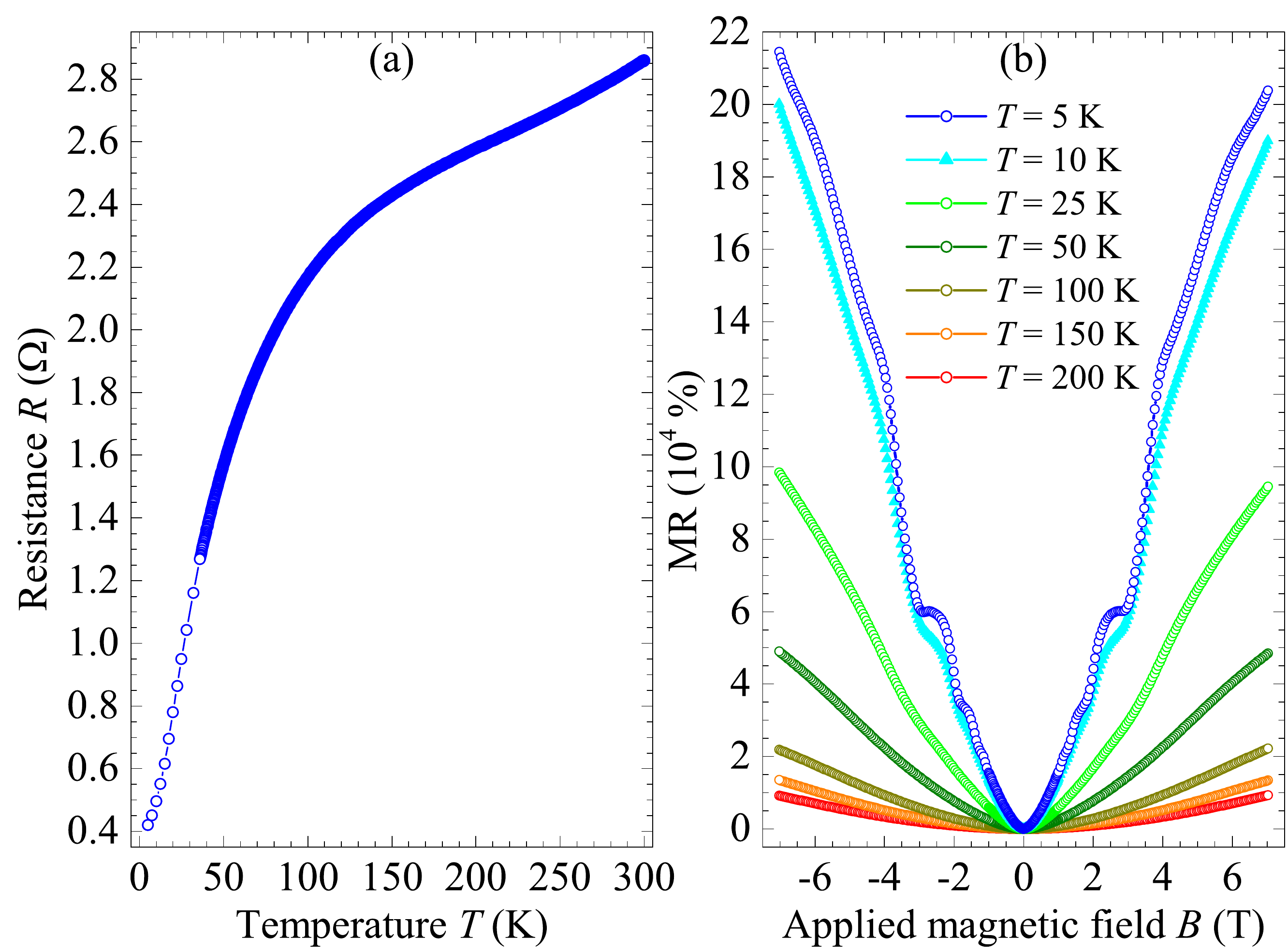}
\caption{Data of sample U6: (a) temperature dependence of the
resistance. (b) The magnetoresistance
        to $B = \pm 7$ T at different temperatures.}
\label{fig:SampleU6}
\end{figure}

\autoref{fig:SampleU6}(a) shows the temperature dependence of
sample U6 of 114~nm thickness. The temperature dependence shows a
metalliclike behavior in all  measured range and a ratio of $R(T =
300 ~\text{K})/R(T = 5 ~\text{K}) \sim 6.80$.
\autoref{fig:SampleU6}(b) shows the MR at different constant
temperatures.

The data of the thicker sample U6 indicate that it has a larger
contribution of interfaces than sample U5, in agreement with
literature.\cite{Zoraghi2018,bar19} The MR in
\autoref{fig:SampleU6} at $T=5$ K shows sharper SdH oscillations
in comparison to those in \autoref{fig:SampleU5}, as expected.


\end{document}